\documentclass{mn2e}
\usepackage{epsfig}
\usepackage{epsf}
\usepackage{graphicx}

\title[SN~2011hw]{SN~2011hw: Helium-Rich Circumstellar Gas and the Luminous
Blue Variable to Wolf-Rayet Transition in Supernova Progenitors}

\author[Smith et al.]{Nathan Smith$^1$\thanks{Email:
    nathans@as.arizona.edu}, Jon C.\ Mauerhan$^1$, Jeffrey M.\
  Silverman$^2$, Mohan \newauthor  Ganeshalingam$^2$, Alexei
  V. Filippenko$^2$, S. Bradley Cenko$^2$, Kelsey I.\ Clubb$^2$,
 \newauthor and Michael Kandrashoff$^2$
 \\ $^1$Steward Observatory, University of
  Arizona, 933 North Cherry  Avenue, Tucson, AZ 85721, USA \\
  $^2$Department of Astronomy, University of California, Berkeley, CA
  94720-3411, USA}

\begin{document}
\date{Accepted 0000, Received 0000, in original form 0000}
\pagerange{\pageref{firstpage}--\pageref{lastpage}} \pubyear{2002}
\def\arcdeg{\degr}
\maketitle
\label{firstpage}

\begin{abstract}

  We present optical photometry and spectroscopy of the peculiar
  Type~IIn/Ibn supernova (SN) 2011hw.  Its light curve exhibits a
  slower decline rate than normal SNe~Ibc, with a peak absolute
  magnitude of $-$19.5 (unfiltered) and a secondary peak 20--30 days
  later of $-$18.3 mag ($R$).  Spectra of SN~2011hw are highly unusual
  compared to normal SN types, most closely resembling the spectra of
  SNe~Ibn. We center our analysis on comparing SN~2011hw to the
  well-studied Type~Ibn SN~2006jc.  While the two SNe have many
  important similarities, the differences are quite telling: compared
  to SN~2006jc, SN~2011hw has weaker He~{\sc i} and Ca~{\sc ii} lines
  and relatively stronger H lines, its light curve exhibits a higher
  visual-wavelength luminosity and slower decline rate, and emission
  lines associated with the progenitor's circumstellar material (CSM)
  are narrower.  One can reproduce the unusual continuum shape of
  SN~2011hw with roughly equal contributions of a 6000~K blackbody and
  a spectrum of SN~2006jc.  We attribute this blackbody-like emission
  component and many other differences between the two SNe to extra
  opacity from a small amount of additional H in SN~2011hw, analogous
  to the small H mass that makes SNe~IIb differ from SNe~Ib.  Slower
  speeds in the CSM and somewhat elevated H content suggest a
  connection between the progenitor of SN~2011hw and the class of
  Ofpe/WN9 stars, which have been associated with luminous blue
  variables (LBVs) in their hot quiescent phases between outbursts,
  and are H-poor --- but not H-free like classical Wolf-Rayet (WR)
  stars.  Thus, we conclude that the similarities and differences
  between SN~2011hw and SN~2006jc can be largely understood if their
  progenitors happened to explode at different points in the
  transitional evolution from an LBV to a WR star.

\end{abstract}


\section{INTRODUCTION}

The diverse class of Type~IIn supernovae (SNe) provides some of our
most puzzling clues about the immediate pre-SN evolution of massive
stars. SNe~IIn are known for their namesake narrow and
intermediate-width hydrogen emission lines that usually dominate the
spectrum (see Filippenko 1997 for a review of SN spectral types), and
which arise from the forward shock of the SN interacting with dense
circumstellar material (CSM). The very dense and massive CSM around
some of the most luminous SNe requires massive episodic shell
ejections shortly before core collapse (see, e.g., Chugai et al.\
2004; Woosley et al.\ 2007; Smith et al.\ 2008a, 2010; van Marle et
al. 2010), the physical parameters of which most closely resemble the
giant eruptions of luminous blue variables (LBVs) like $\eta$ Carinae
(Smith \& Owocki 2006; Smith et al.\ 2003).

Although the idea that stars in the LBV phase could undergo core
collapse was not expected from stellar evolution models, the
LBV/SN~IIn connection inferred from properties of the CSM is
reinforced by the detection of luminous LBV-like progenitors of three
SNe~IIn (Gal-Yam \& Leonard 2009; Smith et al.\ 2011a,b; Kochanek et
al.\ 2011).\footnote{Note, however, that the candidate progenitor of
  SN~2010jl could also be a young star cluster because this SN has not
  yet faded.}  The core-collapse explosion of LBVs violates
expectations of standard stellar evolution models, where massive stars
are supposed to undergo only a very brief (10$^4$--10$^5$ yr)
transitional LBV phase, and then spend 0.5--1 Myr in the core-He
burning Wolf-Rayet (WR) phase before finally exploding as normal
SNe~Ibc (Meynet et al.\ 1994; Heger et al.\ 2003).  In this paper we
discuss a SN whose progenitor star suffered core collapse while
apparently still in the transition from an LBV to a WR star.  Combined
with the LBV-like progenitors of SNe~IIn, this underscores the notion
that stellar evolution models are missing essential aspects of the end
stages of massive stars.

An interesting extension of the Type~IIn phenomenon was most
dramatically illustrated by SN~2006jc (Foley et al.\ 2007; Pastorello
et al.\ 2007; Smith et al.\ 2008b), which exhibited spectral
signatures of strong CSM interaction.  Spectra of SN~2006jc had
relatively narrow lines similar to those of SNe~IIn, but seen mainly
in He~{\sc i} emission lines --- there was only a trace amount of H in
the spectrum.  It is therefore referred to as a ``Type Ibn'' event or
a peculiar Type~Ib, instead of a Type~IIn.  SN~2006jc has been studied
in detail over a wide range of wavelengths (Foley et al.\ 2007;
Pastorello et al.\ 2007, 2008a; Smith et al.\ 2008b; Immler et al.\
2008; Tominaga et al.\ 2008; Nozawa et al.\ 2008; Di Carlo et al.\
2008; Matilla et al.\ 2008; Anupama et al.\ 2009; Sakon et al.\ 2009;
Chugai 2009).  It has important implications for understanding the
broader class of SNe~IIn and Ibn with CSM interaction, because so far
it is the only object that was actually observed to have a
non-terminal LBV-like outburst just 2 yr prior to explosion (Itagaki
et al.\ 2006; Pastorello et al.\ 2007).  Prior to the example of
SN~2006jc, LBV-like eruptions were not known to occur in
H-poor/He-rich stars.  In particular, the very short timescale of only
2 yr confirms conjectures from studies of SNe~IIn that episodic bursts
of mass loss (rather than steady winds) can occur immediately before
core collapse.  If the eruptive events that cause SNe~IIn and Ibn are
indeed synchronised with core collapse, this may provide clues to the
nature of the underlying mechanism (see, e.g., Woosley et al.\ 2007;
Quataert \& Shiode 2012).

While Type~IIn explosions make up 8--9\% of all core-collapse SNe in
the Lick Observatory Supernova Search sample (Smith et al.\ 2011c; Li
et al.\ 2011), the Type~Ibn events like SN~2006jc represent a
substantially smaller fraction.  The ``peculiar'' Type~Ibc supernovae,
which constitute 4\% of the same sample, included some examples of
SN~2006jc-like objects but also had other unusual objects, so the
fraction of SN~2006jc-like events is likely to be around 1\% or less
of all core-collapse SNe.  This agrees with an independent estimate of
the fraction of SN~Ibn events by Pastorello et al.\ (2008a). Given
their rare occurrance, additional examples are valuable to demonstrate
the diversity of the subclass.  Other suggested members of the
Type~Ibn subclass are SN~2002ao, SN~1999cq, and SN~2000er (Matheson et
al.\ 2000; Foley et al.\ 2007; Pastorello et al.\ 2008a).  SN~2005la
had prominent narrow H and He lines, but may be a transitional case
between SNe~IIn and Ibn (Pastorello et al.\ 2008b).\footnote{Of these
  suggested members of the Type~Ibn class, only SN~1999cq and
  SN~2002ao showed an unusual continuum shape and line ratios similar
  to those of SN~2006jc. SN~2000er and SN~2005la have prominent
  He~{\sc i} lines, but these are seen in some SNe~IIn like SN~1988Z
  and SN~2005ip, so it remains unclear if SN~2000er and SN~2005la
  shoud be considered as part of the same class as SN~2006jc.}

\begin{figure}\begin{center}
\includegraphics[width=3.2in]{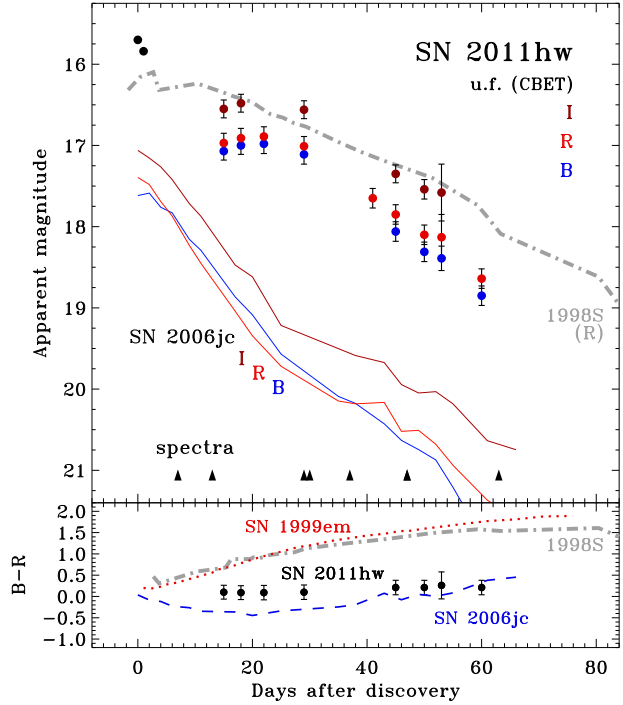}
\end{center}
\caption{(Top:) Optical $B$, $R$, and $I$ photometry of SN~2011hw
  obtained with the 1~m Nickel telescope at Lick Observatory (filled
  dots; see Table 1).  Unfiltered photometry (marked ``u.f.'') from
  early reports (Dintinjana \& Mikuz 2011) is included for comparison;
  these are not the result of our observations.  The thin solid lines
  show the $B$, $R$, and $I$ light curves of SN~2006jc (from Foley et
  al.\ 2007), and the grey dot-dashed line shows the $R$ light curve
  of SN~1998S (Fassia et al.\ 2000), both as they would appear at the
  same distance and with the same extinction as SN~2011hw.  Triangles
  show the dates on which we obtained spectra of SN~2011hw. (Bottom:)
  The observed $B-R$ colour curve of SN~2011hw (solid black dots)
  compared to the same for the SN~Ibn 2006jc (blue dashed line), the
  SN~II-P 1999em (red dotted line; Leonard et al.\ 2002), and the
  SN~IIn 1998S (grey dot-dashed).}\label{fig:phot}
\end{figure}

SN~2011hw provides another case of the Type~Ibn phenomenon.  It was
discovered on 2011 Nov.\ 18.7 (UT dates are used throughout this
paper) by Dintinjana \& Mikuz (2011), occurring in an anonymous host
galaxy whose nucleus was about 8\arcsec\ (3.5--4 kpc) away from the SN
position.  The explosion date is not well constrained by pre-discovery
upper limits, with the most recent pre-discovery upper limit in
December 2010 (Dintinjana \& Mikuz 2011). Valenti et al.\ (2011)
obtained the first spectrum and noted the narrow He~{\sc i} emission
lines. They suggested a similarity to the transitional SN~IIn/Ibn
2005la (Pastorello et al.\ 2008b).  While both SN~2011hw and SN~2005la
do exhibit strong He~{\sc i} emission lines, our spectral comparison
below suggests that SN~2011hw is much more closely related to
SN~2006jc, with only a small increase in the amount of H present in
the CSM.  We suggest that the similarities and differences between
SN~2011hw and SN~2006jc can be understood if their progenitors were at
different points in the transitional evolution from an LBV to a WR
star when they exploded.

\section{OBSERVATIONS}

Following the discovery of SN~2011hw, we obtained $B$, $R$, and
$I$-band photometry using the 1~m Nickel telescope at Lick
Observatory.  The background emission of the rather compact host
galaxy was problematic for simple aperture photometry. This was
particularly true for images obtained under relatively poor seeing
conditions, when the galaxy nucleus and SN image began to partially
overlap, and for the late-time epochs when the SN became faint. Thus,
we employed the IDL Starfinder code to perform point-spread function
(PSF) fitting photometry to extract the SN flux.  For each epoch, a
PSF model was constructed using 5 suitably bright, isolated field
stars. The SN was automatically recovered by the Starfinder code in
all epochs and the galaxy was modeled as local background. The
photometric zero point was derived from the USNO-$B$ photometry of the
PSF stars and two fainter field stars (after converting the magnitude
values to the Johnson system). The photometric uncertainties were
taken as the standard deviation of the SN magnitude values derived by
comparison with each of the individual calibration stars.  Our
resulting $BRI$ magnitudes are listed in Table~\ref{tab:phottab} and
plotted in Figure~\ref{fig:phot}.

\begin{table}\begin{center}\begin{minipage}{3.2in}
      \caption{Lick Photometry of SN~2011hw}  \scriptsize
\begin{tabular}{@{}lcccccc}\hline\hline
  JD        &B &err &R &err &I &err  \\  \hline
            &(mag) &(mag) &(mag) &(mag) &(mag) &(mag)  \\  \hline
 2455898.5    &17.07   &0.11  &16.97   &0.12  &16.55   &0.11 \\
 2455901.5    &17.00   &0.11  &16.91   &0.12  &16.48   &0.11 \\
 2455905.5    &16.98   &0.12  &16.89   &0.12  &...     &...  \\
 2455912.5    &17.11   &0.12  &17.01   &0.12  &16.56   &0.11 \\
 2455924.5    &...     &...   &17.65   &0.12  &...     &...  \\
 2455928.5    &18.06   &0.12  &17.85   &0.12  &17.35   &0.11 \\
 2455933.5    &18.31   &0.12  &18.10   &0.12  &17.54   &0.12 \\
 2455936.5    &18.39   &0.15  &18.13   &0.28  &17.58   &0.35 \\
 2455943.5    &18.85   &0.12  &18.64   &0.12  &...     &0.12 \\
\hline
\end{tabular}\label{tab:phottab}
\end{minipage}\end{center}
\end{table}

\begin{figure*}\begin{center}
\includegraphics[width=6.6in]{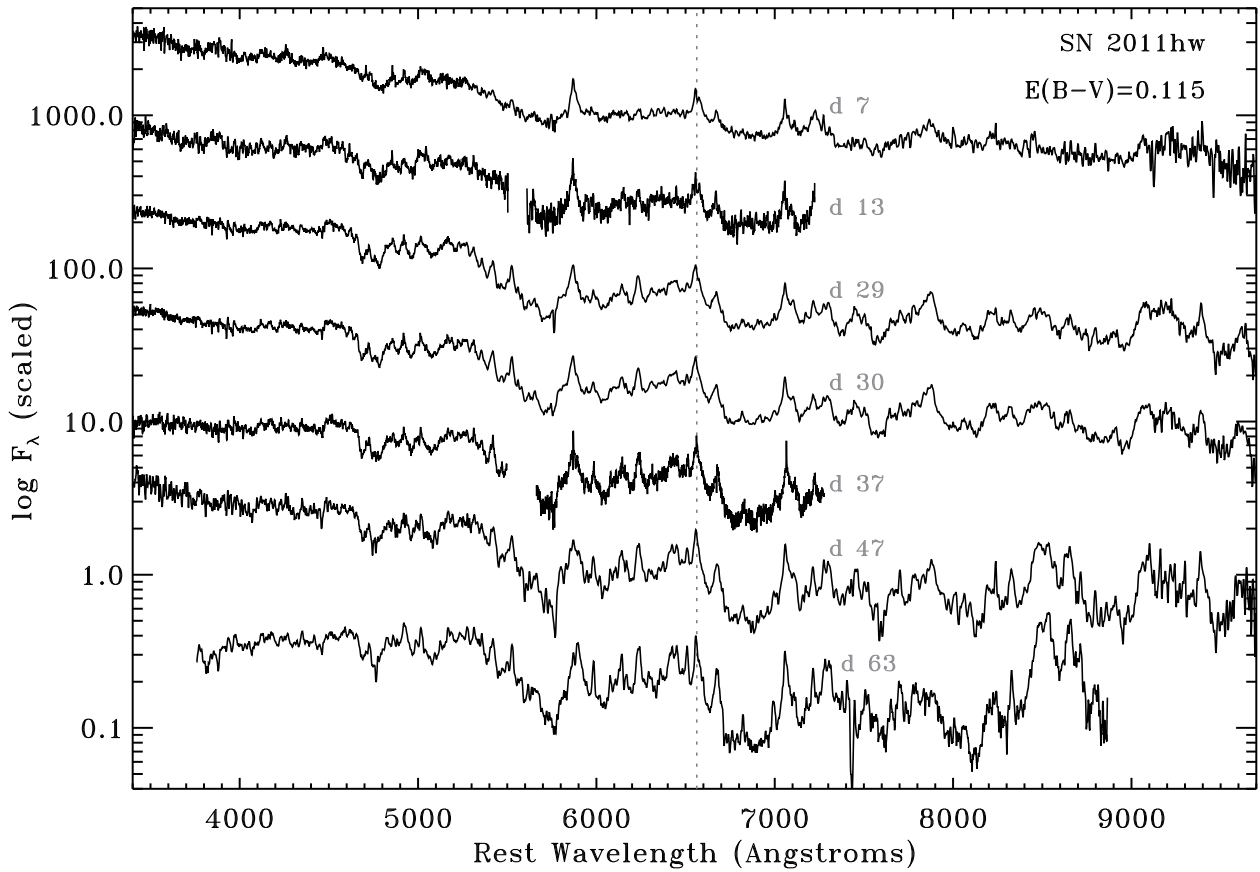}
\end{center}
\caption{Optical spectra of SN~2011hw, with the days after discovery
  indicated (see Table 2).  The vertical dashed line marks the
  wavelength of H$\alpha$.  Note that all spectra have been dereddened
  by the presumed Galactic value of $E(B-V) = 0.115$
  mag.}\label{fig:spec}
\end{figure*}

\begin{table}\begin{center}\begin{minipage}{3.3in}
      \caption{Spectroscopic observations of SN~2011hw}
\scriptsize
\begin{tabular}{@{}llccc}\hline\hline
  Date &Tel./Inst. &Day &Range (\AA) &$\lambda$/$\Delta\lambda$ \\ 
  \hline
2011\,Nov.\,26   &Lick/Kast &7   &3436--9920   &1400, 900  \\
2011\,Dec.\,02   &Keck/LRIS &13  &3362--5630   &1400      \\
2011\,Dec.\,02   &Keck/LRIS &13  &5740--7390   &2200      \\
2011\,Dec.\,18   &Lick/Kast &29  &3436--9920   &1400, 900  \\
2011\,Dec.\,19   &Lick/Kast &30  &3436--9920   &1400, 900  \\
2011\,Dec.\,26   &Keck/LRIS &37  &3362--5630   &1400      \\
2011\,Dec.\,26   &Keck/LRIS &37  &5740--7390   &2200      \\
2012\,Jan.\,03   &Lick/Kast &47  &3436--9920   &1400, 900  \\
2012\,Jan.\,03   &MMT/B.C.  &47  &5550--7500   &4500      \\
2012\,Jan.\,19   &MMT/B.C.  &63  &3820--8998   &500       \\
\hline
\end{tabular}\label{tab:spectab}
\end{minipage}\end{center}
\end{table}

We also began a spectroscopic monitoring campaign using the Kast
spectrograph (Miller \& Stone 1993) mounted on the 3~m Shane reflector
at Lick Observatory, the Low Resolution Imaging Spectrometer (LRIS;
Oke et al.\ 1995) mounted on the Keck~I 10~m telescope, and the
Bluechannel spectrograph on the Multiple Mirror Telescope (MMT). The
spectra were generally obtained at low airmass or with an atmospheric
dispersion corrector; otherwise, the parallactic angle (Filippenko
1982) was used to minimise chromatic differential slit losses. Our
data reduction followed standard techniques as described by Silverman
et al.\ (2012). Our spectroscopic observations are summarised in
Table~\ref{tab:spectab}, and an overview of the spectra is shown in
Figure~\ref{fig:spec}.

Most of the data had moderate resolution ($\lambda / \Delta\lambda
\approx 900$ and 1400 for red and blue sides of the Kast spectra,
respectively) and covered a large wavelength range, but we also
obtained a few epochs with higher resolution ($R \approx 2200$ and
4500 with Keck/LRIS and MMT/Bluechannel, respectively).  These higher
resolution spectra are particularly interesting with regard to the
narrow lines from slow circumstellar gas, as discussed below.

\begin{figure*}\begin{center}
\includegraphics[width=6.6in]{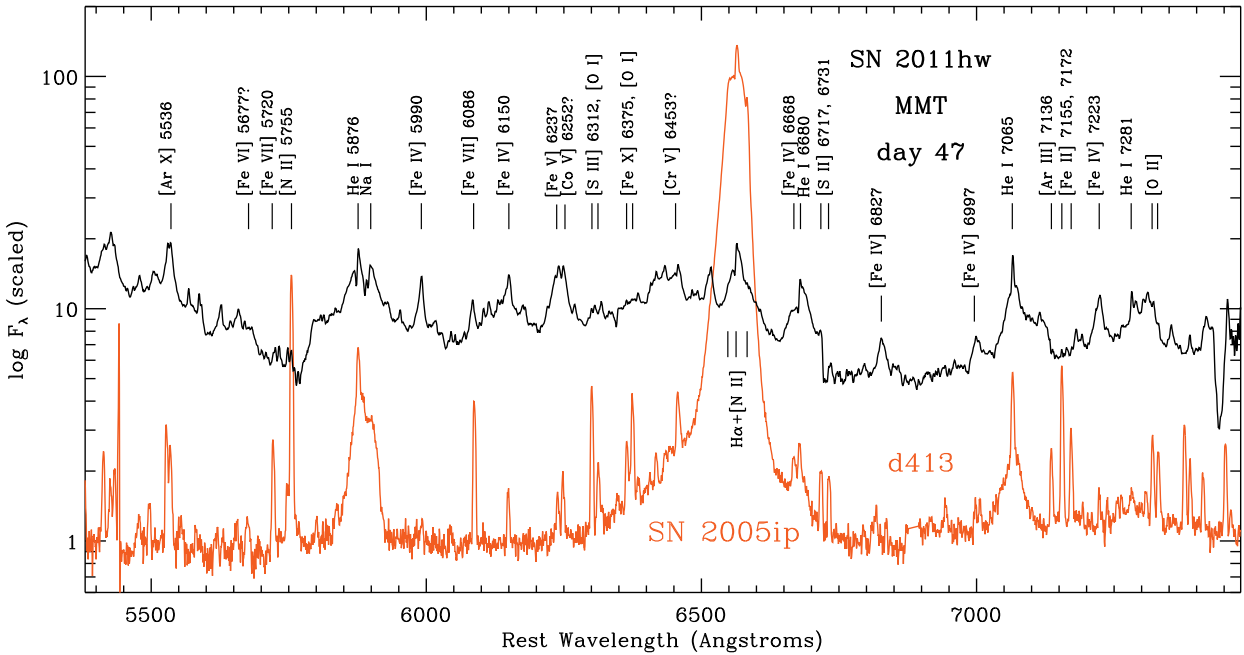}
\end{center}
\caption{High-resolution day 47 MMT spectrum of SN~2011hw, compared to
  a high-resolution spectrum of SN~2005ip from Smith et al.\
  (2009).}
\label{fig:hires}
\end{figure*}

We display the high-resolution day 47 MMT spectrum in
Figure~\ref{fig:hires}. One can detect absorption from the Na~{\sc
  i}~D doublet at the velocity of SN~2011hw, seen as an absorption
notch on the red wing of the He~{\sc i} $\lambda$5876 profile.  This
is not necessarily interstellar-medium absorption that can be
associated with dust, however, since these absorption lines are weaker
or absent in some of our earlier spectra, and there appear to be
Na~{\sc i} D emission components as well (Figures~\ref{fig:spec} and
\ref{fig:hires}).  The resulting P Cygni profile suggests tht the
Na~{\sc i} D is circumstellar, rather than interstellar.  This and the
very blue colour of SN~2011hw are why we do not assume any additional
local reddening and extinction in our analysis below, although it is
difficult to rule out the presence of some additional local dust that
may not be traced by Na~{\sc i} D absorption.

\section{DISCUSSION}

\subsection{Light-Curve Evolution}

As noted in the Introduction, the host of SN~2011hw is an anonymous
galaxy with an unknown distance.  In our spectra, we measure a
redshift of $z = 0.023$ from the narrow-line components in SN~2011hw,
which implies a distance $D = 96.2$ Mpc (assuming H$_0$ = 73 km
s$^{-1}$ Mpc$^{-1}$, $\Omega_{m} = 0.27$, $\Omega_{\Lambda} = 0.73$).
Throughout this paper, we adopt $E(B-V) = 0.115$ mag (Schlegel et al.\
1998) for the value of the line-of-sight Galactic reddening (all
spectra are corrected for this value).  We do not adopt any additional
local extinction, although we cannot rule out the presence of dust
that may be local to SN~2011hw.  With these parameters, the peak
apparent unfiltered magnitude of $\sim 15.7$ at the time of discovery
corresponds to a peak absolute unfiltered (approximately $R$-band)
magnitude of $-$19.5.  This is about 1.5 mag more luminous than
SN~2006jc (Fig.~\ref{fig:phot}; see Foley et al.\ 2007).

The luminosity evolution between discovery and the start of our
photometric monitoring is unclear.  SN~2011hw appears to have declined
quickly from its initial peak, but then leveled off with a plateau or
secondary peak at 20--30 days in our photometry, with an absolute $R$
magnitude of about $-$18.3.  Thereafter, SN~2011hw faded at a
relatively constant rate of $0.050 \pm 0.002$ mag d$^{-1}$ in the $R$
band (Fig.~\ref{fig:phot}), slower than the $0.086 \pm 0.003$ mag
d$^{-1}$ $R$-band decline rate of SN~2006jc before it reached its
radioactive decay tail.  Our photometry ends at about day 60, so we do
not have a reliable estimate of the $^{56}$Co radioactive decay-tail
luminosity.  This does not provide a useful upper limit to the mass of
$^{56}$Ni, since much of the luminosity even at this late phase might
be dominated by CSM interaction.

A possible explanation of the early light curve is that the secondary
peak might be additional luminosity from CSM interaction if the shock
overran a denser portion of its CSM shell, as seen in the late turn-on
of SN~2008iy (Miller et al.\ 2010), although SN~2011hw would be a far
less extreme case of this.  In Figure~\ref{fig:phot} we show that the
light curve of SN~2011hw has a luminosity and decline rate similar to
those of the well-studied SN~IIn 1998S (Fassia et al.\ 2000), in which
CSM interaction with a dense shell was prominent.

Pastorello et al.\ (2008a) infer that SNe~Ibn fade quickly because
they do not have substantial luminosity from CSM interaction.  From
SN~2011hw, we conjecture that this fast fading of other SNe~Ibn may be
the result of very low H abundance, with the lower opacity allowing
more of their CSM interaction luminosity to escape at shorter
wavelengths.  On the other hand, with substantial H opacity, SNe~IIn
have characteristic emitting temperatures around 6000~K and emit much
of their CSM-interaction luminosity at visual wavelengths.  UV
spectroscopy of future examples of SNe~Ibn could therefore be quite
interesting, to test if their lower H abundance allows much of the CSM
interaction luminosity to escape in the UV.

The $B-R$ colour curve of SN~2011hw is similar to that of SN~2006jc,
but is more constant with time, and both are significantly bluer than
the colour curve of a normal SN~II-P represented by SN~1999em or of a
SN~IIn like SN~1998S (Fig.~\ref{fig:phot}).  The colour of SN~2011hw
is in fact remarkably constant during the time period over which we
observed it, rising slowly by only $\la 0.1$ mag, from roughly 0.1 to
0.2 mag in $B-R$ over $\sim 50$ days.  The blue colour is dominated by
the blue/near-UV excess that closely resembles the shape of the
spectrum in SN~2006jc (see \S 3.4).  A similar excess was seen in the
SN~IIn 2005ip, where the narrower lines showed that the blue/near-UV
``continuum'' bump may be due to a blend of many emission lines formed
in the CSM or in the post-shock gas (Smith et al.\ 2009).

In a normal SN~II-P or SN~IIn, the redder colour is determined by the
$\sim 6500$~K continuum that arises from the H-recombination
photosphere.  In SN~2011hw and SN~2006jc, we suspect that the
prominence of the blue bump is due to the lower continuum opacity that
follows from the lower H abundance.  In cases like SN~2005ip where
H$\alpha$ is extremely strong, a lower effective opacity may arise
instead from a highly clumped CSM (see Smith et al.\ 2009; Chugai \&
Danziger 1994).

\begin{figure}\begin{center}
\includegraphics[width=3.4in]{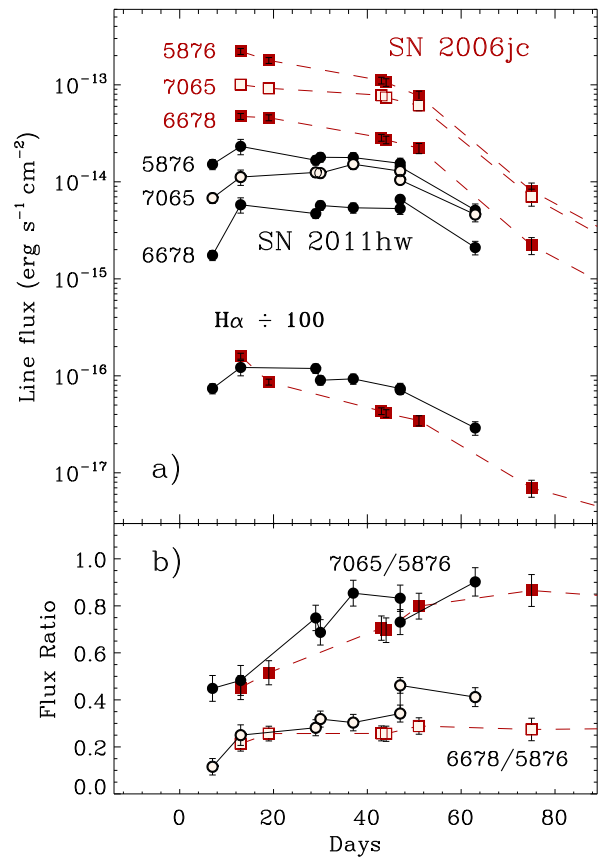}
\end{center}
\caption{(a) Measured line fluxes of He~{\sc i} $\lambda$5876,
  $\lambda$6678, $\lambda$7065, and H$\alpha$ (H$\alpha$ is divided by
  100 for display) from our spectra of SN~2011hw (black dots and solid
  lines) and SN~2006jc (red squares and dashed lines; H$\alpha$ in
  SN~2006jc is also divided by 100). (b) Flux ratios of He~{\sc i}
  lines for the same two SNe.  Data for SN~2006jc are from Smith et
  al.\ (2008b).}\label{fig:lines}
\end{figure}

\subsection{Spectral Evolution}

Overall, one can see from Figure~\ref{fig:spec} that the character of
the spectrum changed very little as SN~2011hw faded by almost 2 mag.
Notable features of the spectrum are the broad blue/near-UV bump in
the continuum (or pseudo continuum), the moderately strong narrow
emission lines of He~{\sc i} and H$\alpha$, and a large number of
weaker narrow (few 10$^2$ km s$^{-1}$) and intermediate-width (few
10$^3$ km s$^{-1}$) emission lines throughout the spectrum.  The
continuum shape and relative strength of He~{\sc i} and H lines
remained largely constant.

One systematic change in our spectra is that as SN~2011hw fades during
the first 60 days, we see a gradual increase in the relative strength
of several narrow and intermediate-width lines
(Figure~\ref{fig:spec}).  A number of narrow lines that are not
present in our first spectrum on day 7 gradually gain in strength at
later times.  Figure~\ref{fig:hires} displays a comparison of our
high-resolution day 47 spectrum of SN~2011hw and a similar spectrum of
SN~2005ip, which reveals that a number of these same narrow lines are
high-ionization coronal lines. While the two spectra have important
differences (especially in H$\alpha$), one can see that several of the
same lines are present in both objects, including lines of [Fe~{\sc
  iv}], [Fe~{\sc v}], [Fe~{\sc vi}], [Fe~{\sc vii}], [Fe~{\sc x}],
[Ar~{\sc x}], etc.  As noted by Smith et al.\ (2009), many of these
lines are commonly seen in active galactic nuclei as well.  This
indicates that as in SN~2005ip, a substantial fraction of X-rays from
the post-shock region are able to penetrate well ahead of the shock
into the CSM.  This may be due either to the lower H opacity,
clumping, or both (see Smith et al.\ 2009).

\begin{figure*}\begin{center}
\includegraphics[width=6.0in]{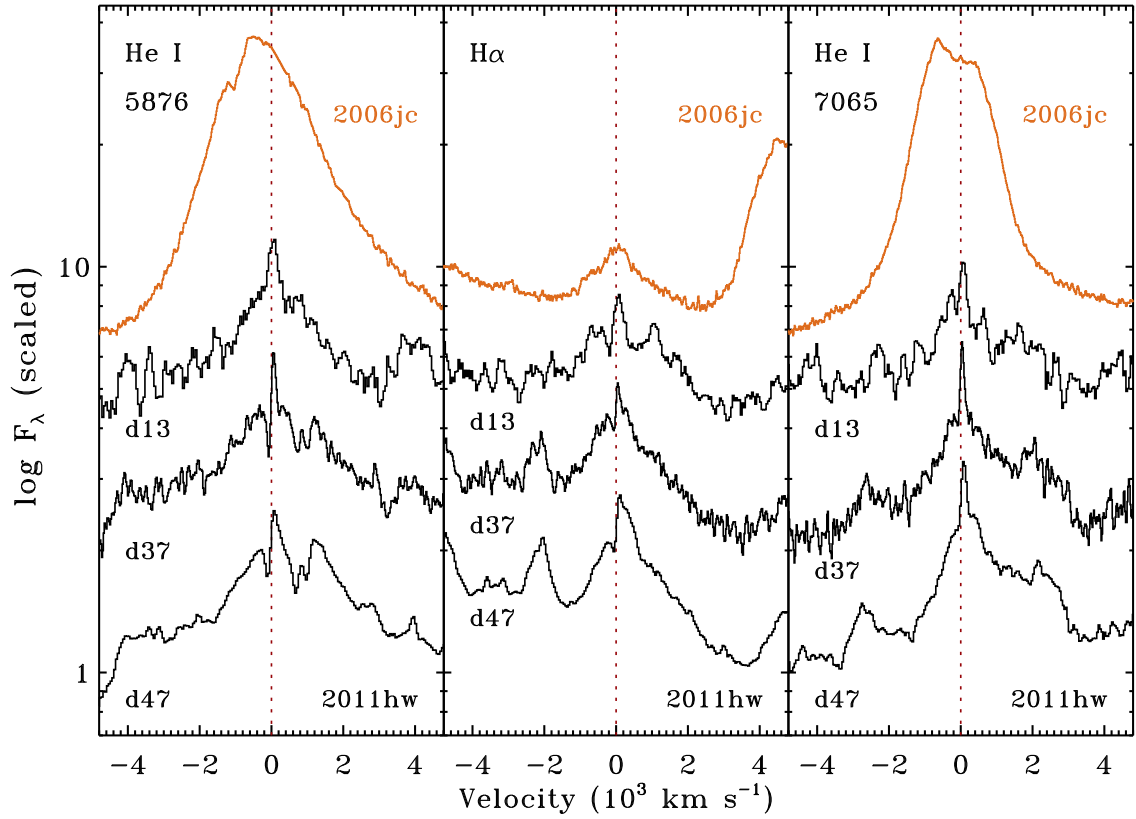}
\end{center}
\caption{The left, middle, and right panels show line profiles of
  He~{\sc i} $\lambda$5876, H$\alpha$, and He~{\sc i} $\lambda$7065,
  respectively, as observed in our higher resolution Keck/LRIS and
  MMT/Bluechannel spectra of SN~2011hw (see Table~\ref{tab:spectab}).
  Spectra of SN~2011hw at days 13, 37, and 47 are in black, while the
  line profiles in a spectrum of SN~2006jc (day 44) are shown in
  orange for comparison.}\label{fig:vel}
\end{figure*}

Figure~\ref{fig:lines} displays the evolution with time of the
measured He~{\sc i} and H$\alpha$ emission-line strengths in
SN~2011hw, as compared to the same lines in SN~2006jc (Smith et al.\
2008b). In general, the observed behaviour is remarkably similar in
the two SNe.  The line flux of H$\alpha$ in the two SNe is nearly
identical, whereas the He~{\sc i} lines are a factor of 5--10 weaker
in SN~2011hw.  This is probably due to a combination of a lower He/H
abundance ratio in SN~2011hw, as well as a radiative transfer effect
wherein a corresponding increase in the H opacity may absorb much of
the harder radiation needed to excite the He~{\sc i} lines.

Aside from this relatively weaker He~{\sc i} emission in SN~2011hw,
the evolution in time of the strength of He~{\sc i} emission lines
(Figure~\ref{fig:lines}a) is very similar in SN~2011hw and SN~2006jc.
Moreover, the evolution in time of the line ratios of He~{\sc i}
$\lambda$7065/$\lambda$5876 and $\lambda$6678/$\lambda$5876
(Figure~\ref{fig:lines}b) is nearly identical for both SNe.  The
He~{\sc i} $\lambda$7065/$\lambda$5876 ratio is particularly
interesting, as ratios rising from $\sim 0.5$ at early times to $\sim
1$ at later times indicate very high electron densities approaching
$10^{10}$ cm$^{-3}$ in the post-shock shell (Almog \& Netzer 1989).
Densities near 10$^{10}$ cm$^{-3}$ represent a critical value for the
nucleation of dust (e.g., Clayton 1979), providing that the gas can
cool sufficiently.  In the case of SN~2006jc, these very high
densities were accompanied by the copious formation of dust grains in
the post-shock shell, indicated by a simultaneous rising near-IR
excess from hot dust, increased optical extinction, and a systematic
blueshift of emission-line profiles (Smith et al.\ 2008b). Although we
lack IR photometry to check for this effect in SN~2011hw, we do not
yet see the systematic blueshift of line profiles in SN~2011hw (see
Figure~\ref{fig:vel}) that would be considered evidence of new dust
forming in the post-shock region.

The reason for the difference may be that at times around day 50 after
explosion (when the density apparently climbs to the critical value
for dust formation in both SNe), SN~2011hw was about 10 times more
luminous than SN~2006jc at the same epoch.  Therefore, while the
carbon-rich gas was able to condense to grains with a temperature of
1700 K in SN~2006jc (Smith et al.\ 2008b), the equilibrium temperature
was probably still too high in the case of SN~2011hw due to its higher
luminosity.  If the density remains high enough for grains to form,
the temperature should become low enough for this to occur at around
day 90--100, assuming that SN~2011hw continues to fade at the same
rate. Because the position of SN~2011hw became too close to the Sun,
we could not continue obtaining spectra after day 63; it will be
interesting to see if the emission-line profiles become asymmetric
when SN~2011hw is observable again.  We note that in the case of
SN~2005ip, which also showed evidence for dust formation despite its
high-excitation spectrum, the IR excess and line-profile evidence of
dust appeared when the SN faded from its main luminosity peak (see
Smith et al.\ 2009; Fox et al.\ 2010).

\subsection{Speeds of the Shocked and Unshocked CSM}

Figure~\ref{fig:vel} shows moderately high-resolution line profiles of
He~{\sc i} $\lambda$5876, H$\alpha$, and He~{\sc i} $\lambda$7065
plotted as a function of velocity at three epochs for SN~2011hw
(black) and SN~2006jc (orange). The full width at half-maximum
intensity (FWHM) values of the intermediate-width component of these
lines are about 1900 km s$^{-1}$ in SN~2011hw, compared to about 3000
km s$^{-1}$ in SN~2006jc.  This component likely represents a slower
speed for the post-shock gas in SN~2011hw, perhaps resulting from a
lower energy explosion or more massive CSM (the hypothesis of more
massive CSM wold be more consistent with the higher luminosity of
SN~2011hw).  The slower shock speed may be partly responsible for the
weaker He~{\sc i} lines in SN~2011hw as compared to SN~2006jc, since
these lines require high excitation.

\begin{figure*}\begin{center}
\includegraphics[width=6.0in]{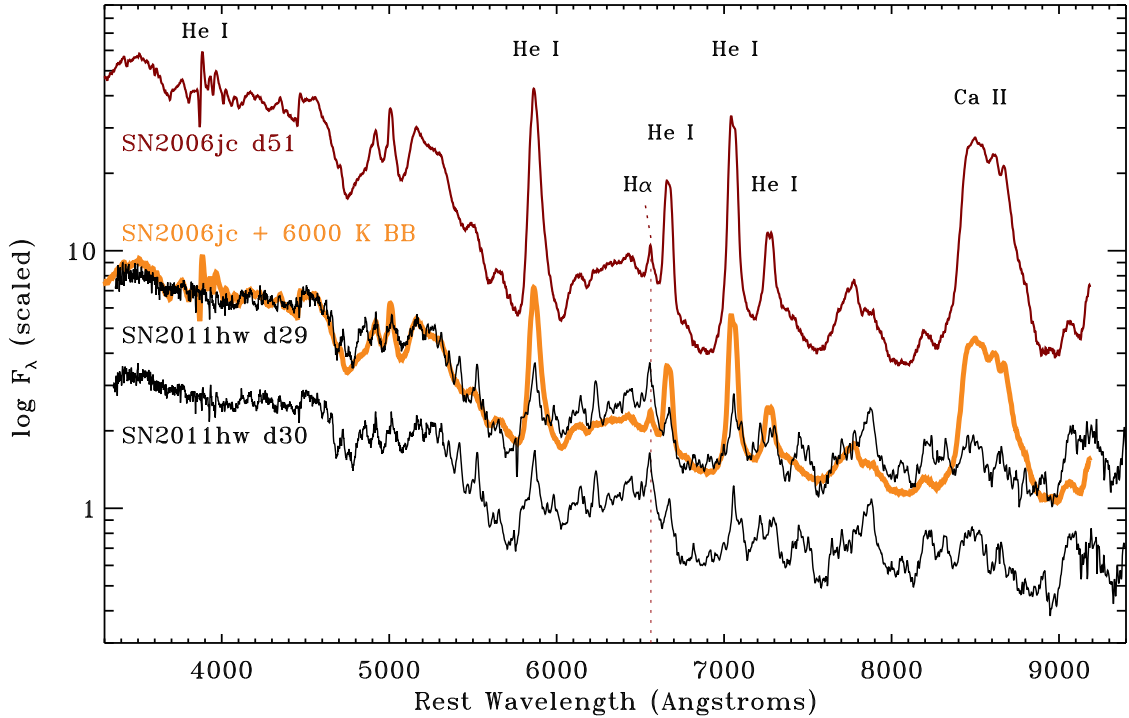}
\end{center}
\caption{Comparison of the optical spectra of SN~2011hw on days 29 and
  30 (black) to the spectrum of SN~2006jc on day 51 (dark red) from
  Smith et al.\ (2008b).  The thick orange spectrum is a composite
  made by adding the SN~2006jc spectrum to a 6000~K blackbody with
  equal contributions (each contributes about half of the red
  continuum flux).}\label{fig:comp}
\end{figure*}

In addition to the slightly narrower width of the intermediate-width
component, SN~2011hw also shows a much narrower emission-line
component that is absent in spectra of SN~2006jc.  Several of these
narrow emission lines exhibit narrow P-Cygni absorption features.
Measured in both He~{\sc i} $\lambda$5876 and H$\alpha$, this
blueshifted P~Cyg absorption trough runs from roughly $-$80 to $-$250
km s$^{-1}$ in the day 47 MMT spectrum (the P-Cyg absorption is not as
prominent in He~{\sc i} $\lambda$7065).  The narrow absorption
velocities appear marginally slower in the day 37 spectrum, although
this spectrum has lower signal to noise ratio. The blue edge of the
P-Cyg absorption therefore indicates a likely maximum speed in the
pre-shock stellar wind of the SN~2011hw progenitor of around 250 km
s$^{-1}$.  In SN~2006jc, the speed of the progenitor's wind was
thought to be a few thousand km s$^{-1}$ (Pastorello et al.\ 2007;
Foley et al.\ 2007; Smith et al.\ 2008a), appropriate for the wind
speed of a compact WR star.  By comparison, the ten-times-slower CSM
speed in SN~2011hw is much slower than expected for a compact massive
star progenitor with a H-depleted, He-rich wind. In other words, the
progenitor of SN~2011hw was probably not a classical WR star.  We
return to this topic in \S 3.6 below.

\subsection{Origin of the  Blue Continuum}

The overall appearance of the optical spectrum of SN~2011hw is
dominated by a large number of weak intermediate-width and narrow
emission lines (discussed above), superposed on a very irregularly
shaped continuum that cannot be reproduced by a combination of
blackbody components.  Qualitatively, the shape of the underlying
continuum of SN~2011hw resembles that of SN~2006jc, except that the
corrugations in the spectrum and the blue excess emission seem more
muted in SN~2011hw.  We therefore experimented with diluting the
spectrum of SN~2006jc using a smooth continuum component to match the
lower contrast of the features in SN~2011hw.  We found that by
diluting the strong features in the spectrum of SN~2006jc with a
6000~K blackbody (equal contributions of each in the red continuum),
we can attain a close match to the shape and strength of the blue
continuum seen in SN~2011hw (Figure~\ref{fig:comp}).  Some emission
features in the SN~2006jc spectrum are still stronger in this
comparison, most notably the narrow He~{\sc i} emission lines and the
Ca~{\sc ii} near-infrared triplet.  The Ca~{\sc ii} triplet
strengthens at later times in SN~2011hw (Figure~\ref{fig:spec}), so
this difference is likely due to different phases in the spectral
evolution.  Ignoring these, the spectra match surprisingly well.

Thus, whatever the origin of the blue/near-UV continuum bump may be in
SN 2011hw, it is likely to have the same origin as in SN~2006jc.
Previously, we speculated that the blue continuum in SN~2006jc was
actually a pseudo-continuum caused by the blending of a large number
of broad and intermediate-width fluorescent emission lines that are
mostly from Fe (Foley et al. 2007; Smith et al.\ 2008b). These lines
produced enhanced blue flux shortward of 4700 \AA, as well as broad
bumps at 4900--5400 \AA\ and at 6100--6600 \AA.  Chugai (2009) showed
that the ``continuum'' of SN~2006jc could indeed be produced by a
large number of blended CSM lines, as postulated earlier.  This idea
received additional support in the case of SN~2005ip (Smith et al.\
2009), which had a similar overall shape for its blue excess emission
--- but in the case of SN~2005ip, the lines were narrower and it was
easier to see that the pseudo-continuum actually broke up into a large
number of emission lines.  As with the presence of narrow
high-excitation coronal lines (discussed above), the blue/near-UV
excess might be a consequence of a lower effective optical depth in
the CSM, allowing more of the hard photons from the post-shock region
to penetrate farther into the CSM.  The lower effective optical depth,
in turn, may be due either to a lower H abundance (compared to normal
SNe~IIn) or to clumping in the CSM, or both.

The continuum component that provides the best match when we dilute
the SN~2006jc spectrum is a 6000~K blackbody, as noted above.
Temperatures lower than 5300~K and above 6800~K yielded significant
discrepancies in the overall continuum shape when compared to
SN~2011hw.  The fact that the best-matching temperature is around
6000~K is significant.  Apparent continuum shapes that match
temperatures of 6000--7000~K are typical in the spectra of SNe~IIn
that are fading after their main luminosity peak (see Smith et al.\
2010).  This is probably due to the temperature of the recombination
photosphere within the cold dense shell of a SN~IIn (Smith et al.\
2008a; Dessart et al. 2009), similar to the apparent temperatures in
normal SNe~II-P (e.g., Dessart \& Hillier 2011).  The temperature of
$\sim 6000$~K therefore suggests that the difference in contrast of
broad continuum features between SN~2011hw and SN~2006jc is the result
of an increase in opacity from a small additional amount of H in the
CSM of SN~2011hw.  The elevated H opacity may also explain why the
Ca~{\sc ii} IR triplet strengthened later (days 47--63) than in the
case of SN~2006jc (already present on day 13), if the Ca~{\sc ii}
emitting region was behind the photosphere until later times. The role
of H abundance is discussed further below.

\begin{figure*}\begin{center}
\includegraphics[width=6.0in]{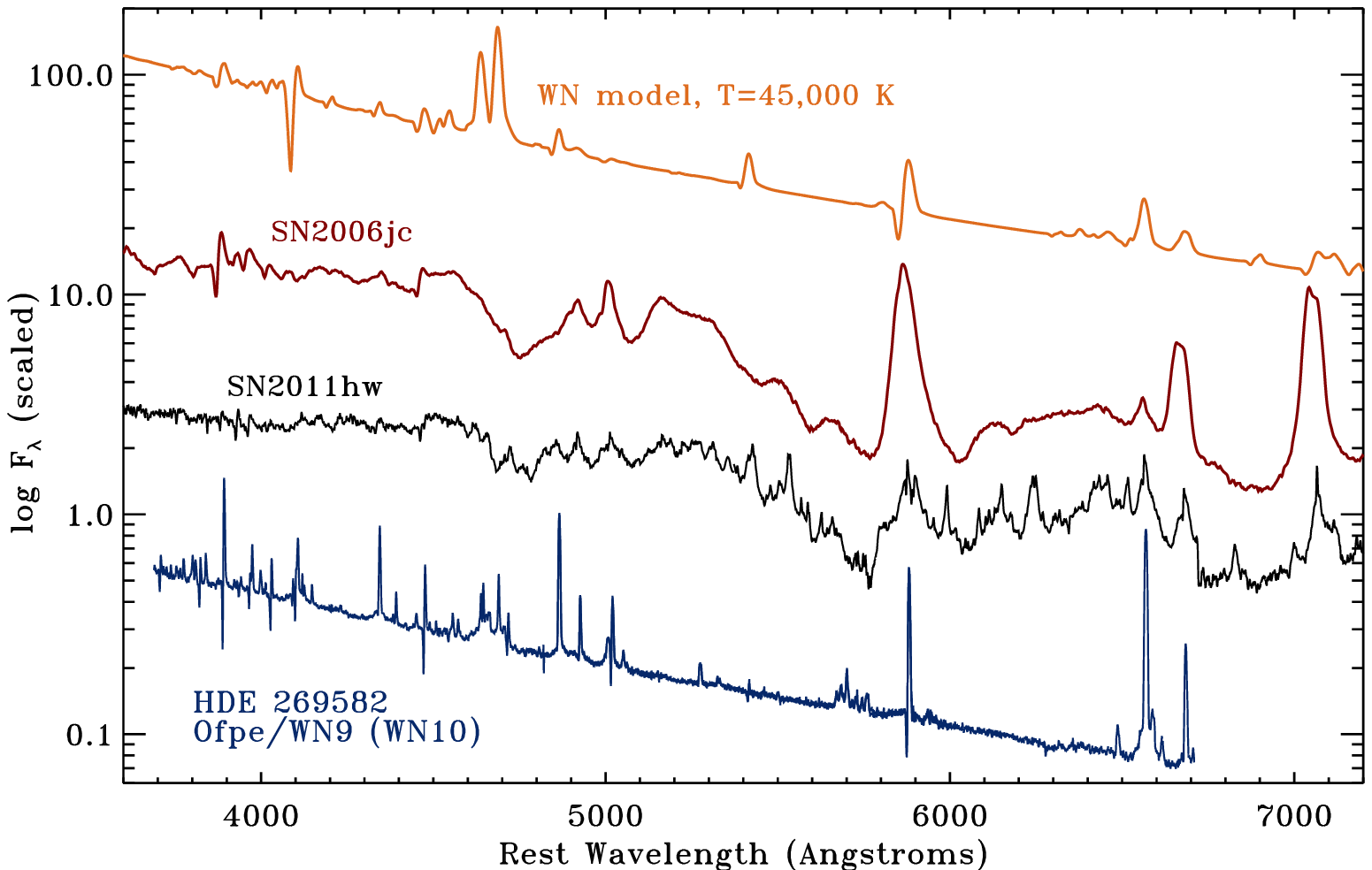}
\end{center}
\caption{Comparison of the spectra of SN~2011hw and SN~2006jc (same as
  in earlier figures) to two types of evolved stars.  At the bottom we
  show the spectrum of the Ofpe/WN9 star HDE~269582 (a typical
  late-type WN star with H, also classified as WN10), and at the top
  we show a model spectrum of an early-type (H-free) WN star.  This
  illustrates the similarity between the narrow lines in the Ofpe/WN9
  star and the narrow CSM lines in SN~2011hw, and also the
  corresponding similarity between the broad-line profiles of
  SN~2006jc and a H-free WR star.  The individual line ratios and
  continua in the spectra obviously differ between the SNe and stellar
  spectra.  The spectrum of HDE~269582 is from the OB Zoo (Walborn \&
  Fitzpatrick 2000; see {\tt http://ftp.astronomy.villanova.edu/}),
  and the WR star is a model spectrum of a He star with $T_{\rm eff}$
  = 45,000~K from the Potsdam grid of WR models (Hamann et al.\
  2004).}
\label{fig:stars}
\end{figure*}

\subsection{Influence of the H Abundance in CSM Interaction}

SN~2006jc was our first well-observed example of a core-collapse SN
interacting with dense H-poor/He-rich CSM (see Foley et al.\ 2007;
Pastorello et al.\ 2007, 2008a; Smith et al.\ 2008b; Immler et al.\
2008), and it showed significant differences in many observed
properties compared to more common cases of H-rich CSM interaction in
SNe~IIn.  SN~2011hw was very similar to SN~2006jc, but appears to have
an elevated H abundance.  The differences between SN~2006jc and
SN~2011hw therefore permit us to understand the impact of adding a
small amount of H to otherwise H-poor CSM being overrun by a SN shock.
Although many apparent consequences of a slightly elevated H abundance
have been discussed in the various sections above, we briefly list
them again here.  This is meant to underscore how the hypothesis of a
slightly elevated H abundance can explain the observed differences
between SN~2011hw and SN~2006jc in a self-consistent way.  For each,
we describe the way in which the H abundance may act to cause the
observed change.

(1) {\it Relatively stronger H$\alpha$ and weaker He~{\sc i} emission
  lines in SN~2011hw.}  The stronger H$\alpha$/He~{\sc i} ratio may
obviously be a direct result of an elevated H abundance, but there is
likely an effect that extra opacity provided by H may provide
additional cooling and may also absorb much of the Lyman continuum
radiation required to excite the He~{\sc i} lines.

(2) {\it The higher luminosity in the secondary peak 20--30 days after
  discovery in SN~2011hw, and the subsequent slower decline rate than
  SN~2006jc.} In \S 3.1 (Figure~\ref{fig:phot}), we discussed how
SN~2011hw has a bump in its light curve 20--30 days after discovery,
and at that time has a substantially higher luminosity than SN~2006jc.
Moreover, it fades from this peak at a slower rate than SN~2006jc
fades from its main peak, comparable to the fading rate of a SN~IIn
like SN~1998s (Figure~\ref{fig:phot}). The stronger continuum opacity
from H may allow a larger resulting CSM-interaction luminosity from
SN~2011hw at visual wavelengths, whereas the lower H abundance in
SN~2006jc may have allowed it to emit a larger fraction of its CSM
luminosity in the UV and X-rays (e.g., Chugai 2009; Immler et al.\
2008).

(3) {\it The 6000~K blackbody component of the continuum in
  SN~2011hw.}  Spectra of SN~2006jc exhibit a strong blue/near-UV
excess and undulations in the continuum throughout the
visual-wavelength range.  Similar features are seen in the spectrum of
SN~2011hw, but their contrast is muted compared to SN~2006jc. In \S
3.5, we showed that the continuum of SN~2011hw could be matched by
using a spectrum of SN~2006jc diluted with a $\sim 6000$ K blackbody,
with roughly equal contributions of both in the red continuum
(Figure~\ref{fig:comp}).  This additional continuum emission component
is closely related to the higher overall optical luminosity in
SN~2011hw, and the 6000~K temperature is an important clue that it is
extra opacity from H that causes this additional contribution.

(4) {\it Narrower CSM lines in SN~2011hw.} It is generally true that
stars with substantial amounts of H present in their outer envelopes
have much larger radii than stars of similar luminosity and mass
having no H. Pertinent examples are that blue supergiants and LBVs
usually have stellar radii $\sim 10$ times larger than H-depleted WR
stars of similar initial mass and luminosity. The larger radius
translates to a lower escape velocity from the star's surface, and
consequently, a lower value for the observed wind speeds for stars
with H envelopes.  Thus, it is likely that the narrower CSM lines in
SN~2011hw as compared to SN~2006jc (\S 3.4) may be attributable to the
progenitor's relatively higher H content. This difference is also
closely related to a different stage of evolution for the two
progenitors (see \S 3.6 below).

(5) {\it Later appearance of the Ca~{\sc ii} IR triplet}.  While
strong and relatively broad emission from the Ca~{\sc ii} IR triplet
was seen in our first (day 13) spectrum of SN~2006jc (Foley et al.\
2007; Smith et al.\ 2008b), these same Ca~{\sc ii} lines did not begin
to strengthen in our spectra of SN~2011hw until day 47.  They became
the strongest lines in the day 63 spectrum.  If these Ca~{\sc ii}
emission lines arise from SN ejecta crossing the reverse shock, they
could have been hidden behind the SN photosphere at early times in
SN~2011hw because the elevated H abundance caused higher optical
depths at larger radii.  In other words, the absence of the Ca~{\sc
  ii} lines at early times is a consequence of the same effect that
causes SN~2011hw to be more luminous and to fade more slowly than
SN~2006jc, as discussed in point (2) above.

All of the observed differences between SN~2011hw and SN~2006jc
described above can be plausibly attributed to a somewhat higher H
abundance in the CSM and progenitor envelope of SN~2011hw. We suspect
that this is indeed the case, but it would be interesting to pursue
this hypothesis wih radiation-hydrodynamic simulations of CSM
interaction using a range of H abundances. Such numerical work is also
needed to provide a quantitative constraint on the actual H/He
abundance ratio. In the absence of such simulations, we draw an
analogy with the differences in SNe~Ib and IIb, suggesting that even a
small amount of H ($\la 0.1$ M$_{\odot}$; Hachinger et al.\ 2012)
could have a large impact on the resulting emission from H-depleted
CSM interaction.

\subsection{The LBV to WR Transition in SN Progenitors}

The observational evidence discussed above establishes that in many
respects SN~2011hw was very similar to SN~2006jc.  Both exhibited
evidence for shock interaction with very dense He-rich CSM that
requires episodic, eruptive mass loss from the progenitor shortly
before core collapse.  We argued that several key differences between
the two can be understood as a result of SN~2011hw having somewhat
more H in its CSM, while still maintaining a low H/He ratio compared
to SNe~IIn.

From our analysis, we find that this body of evidence suggests that
the progenitor of SN~2011hw was probably a late-type WN star with H or
a member of the Ofpe/WN9 class of stars (Bohannan \& Walborn 1989;
Crowther et al.\ 1995; Bianchi et al.\ 2004).  These stars tend to
have slower winds than classical early-type (H-free) WR stars ---
i.e., a few $10^2$ km s$^{-1}$ instead of several $10^3$ km s$^{-1}$
(see Fig.~\ref{fig:stars}).  From studying the host clusters of
Ofpe/WN9 stars, St.-Louis et al.\ (1997) find that they reside among
populations with ages of 3--10 Myr, corresponding to initial masses of
17--100 M$_{\odot}$ if the Ofpe/WN9 stars are poised to explode.

Figure~\ref{fig:stars} compares the spectra of SN~2006jc and SN~2011hw
to spectra of early-type WN and Ofpe/WN9 stars, respectively.  This
latter class of stars\footnote{Note that Smith et al.\ (1994) assigned
  a different spectral class of WN10 or WN11 to this same group of
  stars, instead of the Ofpe/WN9 designation.} is related to the WR
stars in the sense that Ofpe/WN9 and other late-type WR stars with
hydrogen in their spectra are generally thought to represent a
transitional phase between an LBV and a classical H-free WR star; like
LBVs, they are often surrounded by circumstellar shells (Walborn 1982;
Stahl 1987; Smith et al.\ 1994; Pasquali et al.\ 1999).  From the
point of view of expectations from stellar evolution models, it is
therefore quite surprising to see this type of star undergo core
collapse, because the models claim that these stars should just be
starting core-He burning, and should still have another 0.5--1 Myr
left to live.

This contradiction of standard expectations is a close parallel to the
recent recognition that stars that look and behave like LBVs appear to
be exploding to make SNe~IIn, despite theoretical expectations, as
discussed in the introduction.  In fact, these two cases are even more
closely related than this similarity, because there is actually a
direct connection between LBVs and the class of Ofpe/WN9 stars. Based
on classical LBVs like AG~Car and R~127, it has been well established
that some of the more massive LBVs are actually seen to have Ofpe/WN9
spectral types in their hotter quiescent phases between outbursts
(Stahl 1986; Stahl et al.\ 1983).  It is therefore generally thought
that many of the other Ofpe/WN9 stars that have not been observed in
eruption may in fact be quiescent or dormant LBVs that are not
recognised as true LBVs simply because we were not looking at the
right time, or perhaps that they are recent graduates of the LBV phase
(Crowther et al.\ 1995; Bianchi et al.\ 2004; Massey et al.\ 2007).
Correspondingly, Ofpe/WN9 stars (also referred to as ``LBV
candidates'') are often surrounded by massive shell nebulae that
closely resemble LBV nebulae, as noted above.

Associating the progenitor of SN~2011hw with an Ofpe/WN9-like star is
consistent with the higher H content in SN~2011hw, as compared to the
progenitor of SN~2006jc, which seemed to be an early-type WR star with
a faster H-poor wind.  SN~2011hw therefore establishes a bridge
between the H-poor progenitor of SN~2006jc and more H-rich LBVs (and
correspondingly, a likely continuum from SNe~IIn to SNe~Ibn),
justifying earlier suggestions that the eruptive progenitor of
SN~2006jc may have been a star that recently left the LBV phase on the
way to becoming a WR star (Foley et al.\ 2007).  SN~2011hw, by
comparison, was apparently not as far along in this journey when it
underwent core collapse.

There is, of course, no reason to restrict one's consideration of
possible progenitors of SN~2011hw to single stars.  Mass transfer in
massive interacting binaries can also drive the transition from H-rich
to He-rich stellar atmospheres and CSM.  RY~Scuti is an example of a
massive binary caught in this phase (Grundstrom et al.\ 2007).  It is
an O9/B0 supergiant eclipsing binary and is surrounded by a very
dense, compact nebula that is strongly He enriched, but not completely
H free (Smith et al.\ 2002).  Smith et al.\ (2011d) recently showed
that RY~Scuti's surrounding CSM nebula was produced by a series of
episodic mass ejections, similar in principle to those required to
produce the dense CSM around SNe~IIn and Ibn.  RY~Scuti is therefore a
potential example of what a progenitor system of SN~2011hw might look
like.

\smallskip\smallskip\smallskip\smallskip
\noindent {\bf ACKNOWLEDGMENTS}
\smallskip
\footnotesize

We thank Peter Blanchard, Joshua Bloom, B. Y. Choi, Daniel Cohen,
Michelle Mason, Adam Miller, Peter Nugent, and Andrew Wilkins for
assistance with the observations. Some of the data presented herein
were obtained at the W. M. Keck Observatory, which is operated as a
scientific partnership among the California Institute of Technology,
the University of California and the National Aeronautics and Space
Administration; the Observatory was made possible by the generous
financial support of the W. M. Keck Foundation.  The authors wish to
recognise and acknowledge the very significant cultural role and
reverence that the summit of Mauna Kea has always had within the
indigenous Hawaiian community; we are most fortunate to have the
opportunity to conduct observations from this mountain. We thank the
staffs at the Lick, Keck, and MMT Observatories for their help during
the observing runs.  The research of A.V.F.'s group at U.C. Berkeley
is supported by National Science Foundation grant AST-0908886,
NASA/{\it Fermi} grant NNX1OA057G, the TABASGO Foundation, Gary and
Cynthia Bengier, and the Richard and Rhoda Goldman Fund.



\begin{thebibliography}{}
\scriptsize

\bibitem[]{} Almog Y., Netzer H.\ 1989, MNRAS, 238, 57

\bibitem[]{} Anupama G.C., et al.\ 2009, MNRAS, 392, 894

\bibitem[]{} Bianchi L., Bohlin R., Massey P.\ 2004, ApJ, 601, 228

\bibitem[]{} Bohannan B., Walborn N.R.\ 1989, PASP, 101, 520

\bibitem[]{} Chugai N.N.\ 2009, MNRAS, 400, 866

\bibitem[]{} Chugai N.N., Danziger I.J., 1994, MNRAS, 268, 173

\bibitem[]{} Chugai N.N., et al.\ 2004, MNRAS, 352, 1213

\bibitem[]{} Clayton D.D.\ 1979, Ap\&SS, 65, 179

\bibitem[]{} Crowther P.A., Hillier D.J., Smith L.J.\ 1995,
  A\&A, 293, 172

\bibitem[]{} Dessart L., Hillier D.J.\ 2011, MNRAS, 410, 1739

\bibitem[]{} Dessart L., Hillier D.J., Gezari S., Basa S.,
  Matheson T.\ 2009, MNRAS, 394, 21

\bibitem[]{} Di Carlo E., et al.\ 2008, ApJ, 684, 471

\bibitem[]{} Dintinjana B., Mikuz H.\ 2011, CBET, 2906, 1

\bibitem[]{} Fassia A., et al.\ 2000, MNRAS, 318, 1093

\bibitem[]{} Filippenko A.V.\ 1982, PASP, 94, 715

\bibitem[]{} Filippenko A.V.\ 1997, ARAA, 35, 309


\bibitem[]{} Foley R.J., Smith N., Ganeshalingam M., Li W.,
  Chornock R., Filippenko A.V.\ 2007, ApJ, 657, L105

\bibitem[]{} Fox O.D., et al.\ 2010, ApJ, 725, 1768

\bibitem[]{} Gal-Yam A., Leonard D.C.\ 2009, Nature, 458, 865

\bibitem[]{} Grundstrom E.D., et al.\ 2007, ApJ, 667, 505

\bibitem[]{} Hachinger S., et al.\ 2012, submitted (arXiv:1201.1506)

\bibitem[]{} Hamann W.-R., Gr\"afener G.\ 2004, A\&A, 427, 697

\bibitem[]{} Heger A., Fryer C.L., Woosley S.E., Langer N.,
  Hartmann D.H., 2003, ApJ, 591, 288

\bibitem[]{} Immler S., et al.\ 2008, ApJ, 674, L85

\bibitem[]{} Itagaki K., et al.\ 2006, IAUC, 8762, 1

\bibitem[]{} Kochanek C.S., Szczygiel, D.M., Stanek, K.Z. 2011, 
ApJ, 737, 76

\bibitem[]{} Leonard D.C., et al.\ 2002, PASP, 114, 35

\bibitem[]{} Li W., et al.\ 2011, MNRAS, 412, 1441

\bibitem[]{} Massey P., et al.\ 2007, AJ, 134, 2474

\bibitem[]{} Matheson T., et al.\ 2000, AJ, 119, 2303

\bibitem[]{} Matilla S., et al.\ 2008, MNRAS, 398, 114

\bibitem[]{} Meynet G., Maeder A., Schaller G., Shaerer D.,
  Charbonnel C.\ 1994, A\&AS, 103, 97

\bibitem[]{} Miller A.A., et al.\ 2010, MNRAS, 404, 305

\bibitem[]{} Miller J.S., Stone R.P.S.\ 1993, Lick
  Obs. Tech. Rep. 66 (Santa Cruz: Lick Obs.)

\bibitem[]{} Nozawa T., et al.\ 2008, ApJ, 684, 1343

\bibitem[]{} Pasquali A., et al.\ 1999, A\&A, 343, 536

\bibitem[]{} Pastorello A., et al.\ 2007, Nature, 447, 829

\bibitem[]{} Pastorello A., et al.\ 2008a, MNRAS, 389, 113

\bibitem[]{} Pastorello A., et al.\ 2008b, MNRAS, 389, 131

\bibitem[]{} Quataert E., Shiode J.\ 2012, submitted
  (arXiv:1202.5036)

\bibitem[]{} Sakon I., et al.\ 2009, ApJ, 692, 546

\bibitem[]{} Schlegel D.J., Finkbeiner D.P., Davis M.\ 1998,
  ApJ, 500, 525

\bibitem[]{} Silverman J.M., et al.\ 2012, submitted (arXiv:1202.2118)

\bibitem[]{} Smith L.J., Crowther P.A., Prinja R.K.\ 1994, A\&A,
  281, 833

\bibitem[]{} Smith N., Chornock R., Li W., Ganeshalingam M.,
  Silverman J.S., Foley R., Filippenko A.V., Barth A.J.\ 2008a,
  ApJ, 686, 467 

\bibitem[]{} Smith N., Foley R.J., Filippenko A.V.\ 2008b, ApJ, 680, 568

\bibitem[]{} Smith N., Gehrz R.D., Stahl O., Balick B., Kaufer A.\ 
  2002, ApJ, 578, 464


\bibitem[]{} Smith N., Li W., Filippenko A.V., Chornock R.\
  2011c, MNRAS, 412, 1522

\bibitem[]{} Smith N., Li W., Silverman J.M., Ganeshalingam M.,
  Filippenko A.V.\ 2011b, MNRAS, 415, 773

\bibitem[]{} Smith N., Owocki S.P.\ 2006, ApJ, 645, L45

\bibitem[]{} Smith N., et al.\ 2003, AJ, 125, 1458

\bibitem[]{} Smith N., et al.\ 2009, ApJ, 695, 1334  

\bibitem[]{} Smith N., et al.\ 2010, ApJ, 709, 856  

\bibitem[]{} Smith N., et al.\ 2011a, ApJ, 732, 63

\bibitem[]{} Smith N., et al.\ 2011d, MNRAS, 418, 1959

\bibitem[]{} Stahl O.\ 1986, A\&A,164, 321

\bibitem[]{} Stahl O.\ 1987, A\&A, 182, 229

\bibitem[]{} Stahl O., et al.\ 1983, A\&A, 127, 49

\bibitem[]{} St.-Louis N., Turbide L., Moffat A.F.J.\ 1997, in
  Luminous Blue Variables: Massive Stars in Transition, 
  ed. A. Nota, H. Lamers (San Francisco: ASP), 187

\bibitem[]{} Tominaga N., et al.\ 2008, ApJ, 687, 1208

\bibitem[]{} Valenti S., et al.\ 2011, CBET, 2906, 1

\bibitem[]{} van Marle A.J., Smith N., Owocki S.P., van Veelen B.\
  2010, MNRAS, 407, 2305

\bibitem[]{} Walborn N.R.\ 1982, ApJ, 286, 452

\bibitem[]{} Walborn N.R., Fitzpatrick E.L.\ 2000, PASP, 112, 50

\bibitem[]{} Woosley S.E., Blinnikov S., Heger A.\ 2007, Nature,
  450, 390

\end{thebibliography}
\end{document}